\documentclass[10pt,conference,compsocconf]{IEEEtran}
\ifCLASSOPTIONcompsoc
  \usepackage[nocompress]{cite}
\else
  \usepackage{cite}
\fi
\ifCLASSINFOpdf
\else
\fi
\usepackage{amsmath,amssymb}
\usepackage[nolist]{acronym}

\usepackage{balance}

\newcommand{\mrarrow}{\text \textendash \textgreater}

\begin{document}
%
\title{Efficient non-anonymous composition operator for modeling complex dependable systems}

\author{\IEEEauthorblockN{Silvano Chiaradonna, Felicita Di Giandomenico and Giulio Masetti}
  \IEEEauthorblockA{ISTI-CNR, Pisa, Italy\\
    Email: silvano.chiaradonna@isti.cnr.it, felicita.digiandomenico@isti.cnr.it, giulio.masetti@isti.cnr.it}
}


\begin{acronym}[MV\_N1AN2s]
  \acro{EPS}{Electrical Power System}
  \acro{SG}{Smart Grid}
  \acro{SAN}{Stochastic Activity Network}
  \acro{AFI}{Abstract Functional Interface}
\end{acronym}

\maketitle

\begin{abstract}
A new model composer is proposed to automatically
  generate non-anonymous model replicas in the context of performability and dependability
  evaluation. 
It is a state-sharing composer that extends the standard anonymous
replication composer in order to share the state of a 
replica among a set of other specific replicas or among the replica and another
external model. This new composition operator aims to improve expressiveness and performance with respect to the  standard anonymous replicator, namely the one adopted by the M{\"o}bius modeling framework.
\end{abstract}


%
\IEEEpeerreviewmaketitle

\section{Introduction}\label{sec:introduction}

Stochastic model-based evaluation is widely used to analyze the
performability, dependability and performance of complex systems. 
A convenient technique to make model construction easier, and to promote an efficient solution process, 
  is based on the hierarchical submodels composition~\cite{SM91,DKS04}.
  To overcome the heavy and error prone manual definition of many submodels, automated procedures are currently available to assist the modeller in generating the hierarchical composed model representing the
  specific instance of the system. A general approach is based on the definition of primitives,
  such as the \textit{Rep} operator proposed in the M{\"o}bius framework~\cite{DCCDDDSW02}, which can be used to represent different components, behaviors or
  characteristics of a system, automatically replicating a single template
  model. Each replica is indistinguishable by other replicas and then it is anonymous. Although satisfactory in general terms, such \textit{Rep} operator shows limitations from the efficiency point of view when employed in the modeling of systems composed of a (large) population of similar, but non anonymous, components. 
  
  In this paper, a new composition operator, called \textit{NARep}, is introduced to deal with efficient non-anonymous replication inside the M\"obius framework. 
  The new operator characteristics are identified and presented in the
  context of performability and dependability
  evaluation.

  \section{The context}\label{sec:context}
  
 The M\"obius modeling framework~\cite{DCCDDDSW02} is a powerful modular environment that supports multiple modeling formalisms and solution techniques.
  Among the others, it supports the \ac{SAN} formalism,
  composition operators \textit{Rep} and \textit{Join}~\cite{SM91}, performance variables
  based on rate and impulse rewards (used also to define performability and dependability measures), analytical and simulative solvers. 

  \textit{Join} operator is a state-sharing composer that 
  brings together two or more submodels.
  \textit{Rep} operator is a special
  case of \textit{Join} that constructs a model consisting of identical copies (replicas) of a submodel.
  \textit{Rep} and \textit{Join} operators are defined at level of \ac{AFI}~\cite{D00,DCCDDDSW02}, 
  a common interface between model formalisms and solvers
  that allows formalism-to-formalism and formalism-to-solver 
  interactions. 
  At \ac{AFI} level, places and activities
  (transitions) of \ac{SAN} correspond to state variables and actions.
  In the following, for sake of simplicity, we will use \ac{SAN}
  models and \ac{SAN}
  notation (places and activities) also to describe \textit{Rep} and \textit{Join} operators.
  The state of a model is represented by the current value of all
  the places defined in the model. 
  A place of a replicated \ac{SAN} model can be local (assume, at the same time, different values in different replicas)
  or shared (same place, with only one value, that can be read or updated by each replica). 
  The \textit{Rep} operator can be used to represent multiple components, behaviors or
  characteristics of a system having identical definition and
  parameters.

  \subsection{Non-anonymous replication in M{\"o}bius}
  
 Non-anonymous replicas of a single \ac{SAN} model can be
 automatically generated through \textit{Rep} operator via a
 sophisticated use of local places and immediate activities as shown
 in~\cite{CDGM14}.
  Local places, representing replica indexes, are set up so that each replica becomes self-aware of its index.
  The specific state, number of ``tokens'' or ``marks'' in Petri nets dialects, of the $i$-th non-anonymous
  replica is represented by the $i$-th entry $P\mrarrow
  Index(i)\mrarrow Mark()$ of a
  shared array-type extended place $P$, i.e., an 
  extended place defined as an array of states, that is shared among \textit{all} the
  replicas. Thus, each replica has read and write access to the state
  of all the other replicas of the same \ac{SAN}.
  In addition, a different submodel that shares the place $P$ with a \textit{Rep} has read and write access to
  the state of each specific replica. 

This approach allows to cope with dependability analysis of complex critical
systems, with a variety of interconnected components organized according to sophisticated topologies, such as electrical or transportation systems.
In~\cite{CDGM16}, which focuses on Smart Grids,  a 
  stochastic modeling framework has been 
  proposed, which uses the \textit{Rep} operator as described above to quantitatively assess representative indicators of the resilience
  and quality of service.
  



  \subsection{Motivations for an enhanced replication operator}

  Using an anonymous replicator to produce non-anonymous replicas
  sounds unnatural, while on the other hand the \textit{Join} operator is impractical for large models.  Moreover, dependencies introduced by the $n$-dimensional array-type extended
  places shared among all the $n$ replicas of a \ac{SAN} model can
  lead to a large time overhead when, during the initialisation phase
  of the simulator, connectivity lists~\cite{W98} are 
  generated.
  Implementing a non-anonymous replication with the \textit{Rep}
  operator and with $n$-dimensional array-type extended places, shared
  among $n$ replicas, slows down the connectivity lists
  construction. In fact, $n^{2}$ checks are needed because each replica can
    access to the state of \textit{all} the other replicas.
  With simple toy models, varying $n$ from $10$ to $50$ the connectivity lists generation time overhead remains acceptable, but varying $n$ from 
  $100$ to $500$ the time overhead can increase of about $100$ times. Following these considerations, we propose the \textit{NARep} (\textit{non-anonymous replication}) new operator. 

  \section{\textit{NARep} definition}\label{sec:NARepDef}
  
  \textit{NARep} automatically constructs a model
  consisting of non-anonymous indexed replicas of a given submodel.
  The index of a specific replica can be accessed from the generic \ac{SAN}
  submodel using a new reserved keyword $repindex()$, corresponding to a method of the C++ \ac{SAN} class.
  Each place $P$ included in a non-anonymous
  model $aSAN$, replicated $n$ times,
  corresponds to $n$
  different place replicas via the \textit{NARep} operator, and place
  replicas are accessed with $aSAN\mrarrow Index(0)\mrarrow P$,\dots,
  $aSAN\mrarrow Index(n-1)\mrarrow P$. The \textit{NARep} operator can
  be used to define whether the 
  replicas of each place $P$ are:  
\begin{enumerate}[(i)]
\item local (default): read and write access to replica place
  $aSAN\mrarrow Index(i)\mrarrow P$ limited only to the
  corresponding $i$-th replica of the \ac{SAN} submodel. The generic
  model $aSAN$, representing the generic replica $repindex()$, can
  access to the corresponding replica $repindex()$ of the place $P$,
  using: $aSAN\mrarrow P\mrarrow Mark()$, or equivalently
  $aSAN\mrarrow Index(repindex())\mrarrow P\mrarrow Mark()$. 
\item place-shared: several place replicas correspond to a single
  place. For example, if $aSAN\mrarrow Index(0)\mrarrow P\mrarrow
  Mark()$ is equal to $aSAN\mrarrow Index(1)\mrarrow P\mrarrow
  Mark()$, then the command $aSAN\mrarrow Index(repindex())\mrarrow P$
  is referred to the same place replica each time that $repindex()$ is
  equal to $0$ or $1$. 
\item rep-shared: the same place replica is shared among several
  replicas of the \ac{SAN} submodel. For example, the place replica
  $aSAN\mrarrow Index(i)\mrarrow P$ can be rep-shared among the
  $(i-1)$-th and $(i+1)$-th replicas of $aSAN$. In this case, the
  generic model $aSAN$ can access to the rep-share places using
  $aSAN\mrarrow Index((repindex()-1)\%n)\mrarrow P$, $aSAN\mrarrow
  Index(repindex())\mrarrow P$ and $aSAN\mrarrow
  Index((repindex()+1)\%n)\mrarrow P$. The
  list of places that are rep-shared with a replica is
  obtained using the command $P\mrarrow repshared()$. Using this command, a generic \ac{SAN} model can
  represent the access of the $i$-th replica only to those components
  that are connected to the $i$-th replica. 
\item up-shared: a local, place-shared or rep-shared place replica can
  be shared among other submodels (not $aSAN$) at the level on top of
  $NARep$. For example, $aSAN\mrarrow Index(0)\mrarrow P$ and
  $aSAN\mrarrow Index(1)\mrarrow P$ can be local and up-shared with
  the place $otherSAN\mrarrow Q$, where the model $otherSAN$ is
  composed with \textit{NARep}. In this case, the $Q$ place included
  in $otherSAN$ is an extended place defined as a $2$ dimensional array,
  each entry having the same type of the place $P$, such that $aSAN\mrarrow
  Index(0)\mrarrow P$ is equal to $Q\mrarrow Index(0)$ and
  $aSAN\mrarrow Index(1)\mrarrow P$ is equal to $Q\mrarrow Index(1)$. 
\end{enumerate}


  \section{Discussion}\label{sec:Discussion}
  
  The proposed \textit{NARep} operator has the following advantages:
  \begin{itemize}
  \item Automatic generation of non-anonymous indexed
    place and \ac{SAN}
    replicas transparent to the modeler; this feature enlarges the framework expressive power.
  \item A significant reduction of the connectivity lists generation
    time overhead can be obtained exploiting only actually existing dependencies instead of the
    assumed complete graph of dependencies. Therefore, the $n^{2}$ checks currently performed by the \textit{Rep} operator constitute the worst case for \textit{NARep}, when full connection is considered.
  \item Symmetries exploitation via the state-space lumping~\cite{DKS04} remains possible.
  \end{itemize}

  Future work will include:
  \begin{itemize}
  \item \textit{NARep} formalism definition, 
  \item integrate \textit{NARep} into M{\"o}bius framework,
  \item evaluate the actual impact of \textit{NARep} on the
    connectivity lists generation time overhead; 
    in particular, a significant improvement is expected for the Smart Grid model
    presented in~\cite{CDGM16}, being the
    electrical and communications networks sparse.
  \end{itemize}

\bibliographystyle{IEEEtran}
\bibliography{EDCC2016}
%

\end{document}